\documentstyle[12pt]{article}
\def\be{\nopagebreak[3]\begin{equation}}
\def\ee{\end{equation}}
\def\ba{\nopagebreak[3]\begin{eqnarray}}
\def\ea{\end{eqnarray}}
\def\nl{\nonumber \\}
        \def\ni{\noindent}
        
\begin{document}
	\begin{center}
        \centerline{}
        \vspace{.5cm}
	{\LARGE\bf Gravity and Geometric Phases\\
         }
	\vspace{0.7cm}
	{\large\em A. Corichi and M. Pierri\\}
	\vspace{0.5cm}
	{ Center for Gravitational Physics and Geometry \\
	Department of Physics, The Pennsylvania State University  \\
	University Park, PA 16802\\ }
	\vspace{0.3cm}
	{\small December 1st, 1994\\ }
	\vspace{0.5cm}
	\end{center}

\begin{abstract}

The behavior of a quantum test particle satisfying the Klein-Gordon equation
in a certain class of 4 dimensional stationary space-times
 is examined. In a space-time  of a spinning cosmic string,
the wave function of a particle in a box is shown to acquire  a
geometric phase when the box is transported around a closed path
surrounding the string.
When interpreted as an Aharonov-Anandan geometric phase,
 the effect is shown to be related to the Aharonov-Bohm effect.
\end{abstract}

\section{Introduction}

The term ``gravitational Aharonov-Bohm effect'' (gAB) has been used
to describe a variety of systems with diverse properties
  \cite{{dowker},{bez}}. Usually it
implies a  behavior different from the one in
Minkowski space-time even
when the motion is restricted to regions in which the space-time is
(locally) flat.
At the classical level,
the parallel transport of vectors and spinors round a cosmic string, for
example,
is nontrivial when the mass $M$ of the string is different from zero.  When
the angular momentum $J$ is non vanishing, the time required for null
(light) rays to travel around the cosmic string depends on the direction
of motion  \cite{ash75a}.
For a quantum test (Klein Gordon) particle  the effect is
manifested in different ways depending on the situation. When bound states
are studied \cite{bez} it is seen that the energy spectrum is modified
depending on the mass and angular momentum of the string. In the
scattering of particles, the scattering amplitude is non trivial and
the phase shifts are modified also depending on $M$ and $J$ \cite{jack89}.

It has also been noted that in linearized gravity, i.e. with $g_{ab}=
\eta _{ab} + h_{ab}$ where $h_{ab}$ is small,
the deparametrized square-root
Hamiltonian of a slowly moving particle
takes the form: $H = (p_i - A_i)^2 + \frac{1}{2} m h_{00}$ where
$A_i= \frac{1}{2} h_{0i}$. $A_i$
transforms like a $U(1)$ connection under infinitesimal
diffeomorphism $x^i \rightarrow x^i + \zeta ^i$ and if
$\partial _{[i} A_{j]} = 0$, the Schr\"odinger equation looks like the
one  in the AB case, $h_{0i}$ playing the role of
the electromagnetic potential \cite{dewitt66}.
Using this analogy, Anandan obtained a geometric phase for a particle
satisfying the Schr\"odinger equation in the  weak field region around a
cosmic string by interfering beams passing on opposite sides
of the string. The wave function thus constructed is, however, not
single-valued \cite{anandan91}.

We will show that this last construction can be extended to
general metrics (without the  weak field assumption) and for the
fully relativistic Klein-Gordon equation (so that the velocity is not
necessarily
small) using only single-valued solutions.
The $g_{0i}$ part of the metric still behaves like a potential.
 Both the Hamiltonian constraint and the Klein Gordon equation
have a very appealing form in terms of a fiducial static metric
on which the potential $A_i$ ``lives''. This will allow us to
identify $g_{0i}$ as a vector potential and, in the case of locally
flat space-times (which include cosmic strings ), discuss a fully relativistic
gAB effect.

The similarities between charged quantum particles in the presence of a
long thin solenoid and scalar quantum particles in the space-time
generated by a cosmic string, are at two levels: 1) The Schr\"odinger
equation for the charged particle and the Klein-Gordon equation
have a similar form ;
2) There exists a ``vector potential'' in the linearized case that
behaves like a electromagnetic $U(1)$ connection. Nevertheless, it
is not apriori clear to what extent $g_{0i}$ can be taken as a genuine
connection
and what its role is in the gAB effect, as compared to the role of $A_i$ in
the electromagnetic AB effect. The aim of this work is to try to clarify
this point.

The structure of the paper is as follows. In section 2, we consider
the Klein-Gordon equation
for a specific class of stationary space-times that can be decomposed into a
fiducial static background metric plus a ``potential $A_i$ term''.
We show that the solutions can be constructed from solutions on the
 ``background'' metric using a prescription introduced by Dirac.
For the specific case of the space-time of a spinning cosmic string further
properties  hint at a gAB effect.
In section 3  we review the electromagnetic Aharonov-Bohm effect.
Following Berry's suggestion, it can be interpreted as an Aharonov-Anandan
 geometric phase i.e. the
holonomy of a natural connection in the Hilbert space of states.
Section 4 recalls that, for a stationary space-time, the space of solutions of
the Klein-Gordon equation can be given the structure of a complex Hermitian
 Hilbert space.
In section 5, we combine the results of sections 2, 3 and 4 to conclude
 that a geometric phase does exist for the spinning cosmic
string. The phase is explicitly constructed. It provides a
fully relativistic gravitational Aharonov-Bohm effect.

Throughout the paper $G= \hbar = c =1 $ units are assumed.

\section{Test Particle on Stationary Space-times}

Stationary solutions of vacuum Einstein equations can describe the
space-time geometry outside rotating bodies. As is well known such metrics
are characterized by having a timelike Killing vector field $t^a$. Let
 us consider the following stationary line-element \cite{maz86}:

\be
 ds^2 = -V^2 (dt- A_i dx^i)^2 + h_{ij} dx^i dx^j \, ,\label{linel}
\ee

\ni
where $V$, $A_i$, $h_{ij}$ are functions on a Cauchy surface $\Sigma$
coordinatized by $x^i$, $i=1,2,3$.
$A_i$ has vanishing
curl $\partial_{[i}A_{j]}=0$.
The time coordinate
$t$ is the affine parameter along
the timelike Killing vector field $t^a$. The freedom left in the the choice of
$t$
 is $t\longrightarrow  t +\Lambda(x^i)$ whereas $x^i$
can be
transformed among themselves completely arbitrarily without changing the
general form of the metric. Under this coordinate
transformation, the object $A_i$ transforms as
$A_i \longrightarrow A_i + \nabla_i \Lambda $.
Under diffeomorphisms of the 3 manifold $\Sigma$,  $A_i$
transforms as a covector. We can therefore interpret $A_i$ as  a $U(1)$
connection under gauge transformations, living on the fiducial
static space-time with metric $\stackrel{\circ}{g}_{ab} = -V^2 \nabla_a t
\nabla_b t + h_{ij}\nabla_a x^i \nabla_b x^j$.

Assume that we have massive scalar quantum test particles
whose behavior is determined by the
Klein-Gordon equation,
\be
(\Box  - m^2) \Psi = 0 \, .\label{kgeq}
\ee
For the line element (\ref{linel}) it takes the form,

\ba
\{ \stackrel{\circ}{\Box} + h^{ij} A_i A_j {\partial_0}^2 & + &
 2\,h^{ij}  A_j \partial_0 \partial_i +
     \nonumber \\
  & + &  \frac{1}{V \sqrt h} \partial_i \left( h^{ij} A_j V {\sqrt h}
 \partial_0 \right)  - m^2  \} \Psi = 0 \, , \label{kgeq2}
\ea

\ni
where $\stackrel{\circ}{\Box}$ is the operator corresponding to
$\stackrel{\circ}{g}_{ab}$,
$h^{ij}$ is the inverse of $h_{ij}$ and $h=\det (h_{ij})$.
Since the background is time independent, we can look for
 solutions of the form

\be \Psi(t,x) = e^{-iEt} \Phi (x) \label{3}\ee

Substituting (\ref{3}) into (\ref{kgeq2}) we obtain
an equation for $\Phi (x)$. It is a curious fact that
its solution can be written in terms of the solution $\Phi_0(x)$ to the
Klein-Gordon
equation $(\stackrel{\circ}{\Box} - m^2 ) \Phi_0 = 0$ on the background
metric $\stackrel{\circ}{g}_{ab}$,

\be
 \Phi (x) = e^{iE\,\int_{\bf x_0}^{\bf x} A_i dx^i} \Phi_0 (x) \label{4}
\ee
This  is the Dirac phase factor method.
Although $\Psi(t,x)$ is a solution, it is multi-valued and therefore
 should be treated with care as we will see
in section 5. For a non-contractible  closed path $\sl C$ the multi-valued
feature of solution (\ref{3}) is transparent.

One remark is in order. The requirement that $\partial_{[i} A_{j]}=0$
means that, locally, the connection $A_i$ is pure gauge, i.e. that locally
 the metric
(\ref{linel}) can be put in a  static form. More precisely, it  implies that
the
twist of the Killing field $t^a$ is zero and therefore the nonintegrabillity
of the
hypersurfaces has a topological origin\footnote{If we regard $F_{ab}=
\nabla_a t_b$ as an ``Electromagnetic two form", its magnetic field
(proportional to the
twist) would be zero; formally the situation is the same as
 in the electromagnetic AB
case.}, namely  $H^1(\Sigma)\neq 0$.

A particular case of (\ref{linel}) is the space-time generated by a spinning
cosmic string, on which we will focus from now on. In cylindrical
coordinates
$(t,\rho ,\theta ,z)$ the line element is given by  (\ref{linel}) with

\ba
V^2 & = & 1 \nonumber \\
A_i & = & - 4 J \nabla_i \theta \nonumber \\
h_{ij} &  = &  \nabla_i \rho \nabla_j \rho + (\alpha \rho)^2 \nabla_i \theta
\nabla_j \theta + \nabla_i z \nabla_j z \, , \label{4a}
\ea

\ni
where $J$ is the angular momentum of the string and $\alpha= 1 - 4\mu$
with $\mu$ the linear mass density. This space-time is flat (outside the
origin) and can be locally written as the Minkowski line element.
As a consequence of non vanishing $\mu$ the spatial slices (t=constant)
are planes with a wedge of angle $\beta = 8 \pi \mu$ removed and edges
 identified, i.e., a cone. For a non vanishing $J$, the
hypersurfaces which are locally orthogonal to the timelike Killing vector
field, can not be globally integrated.

If one takes the Killing vector field $(\frac {\partial}{\partial z})^a$ and
reduces along its orbits, one gets a $2+1$ space-time with a massive spinning
particle at the origin. The mass of the particle is $\mu$ and its angular
momentum is $J$. For a detailed description of such space-time see
 \cite{deser84}. Therefore there exist two equivalent descriptions, one four
dimensional and the other three dimensional.

There are some surprising  results related to this space-time
already in classical physics that are worth pointing out.
Let us examine the effects of parallel transport
of spinors and vectors along non-contractible loops. Although this space-time
is locally flat everywhere (except at the origin) holonomies are non
trivial. The components of vectors undergo a change which is sensitive to
the angular defect $\beta$ as well as the angular momentum
$J$. On the other hand the components of spinors detect just the angular
defect $\beta$.

Finally, let us consider quantum particles. Using (\ref{3}) and
(\ref{4}), it follows that
the solution to the Klein-Gordon equation in this background is given by

\be
\Psi (t,\rho, \theta, z) = e^{-iEt} \,\Phi (\rho, \theta, z) \, ,
\label{5}
\ee

\ni
where

\be
\Phi(\rho, \theta, z) = e^{-i4JE\int_{\theta_0}^{\theta}d\theta}
\,\Phi_0 (\rho, \theta, z) \, ,\label{5a}
\ee
\ni
$\Phi_0(x)$ being a solution on the space-time of a non rotating string.
In the next section we will see how the phase-factor gives rise to some
interesting results.

 \section{Geometric phase and the Aharonov-Bohm effect}

The property that the solution of a ``complicated'' equation can be found
from solutions to a ``simpler'' equation using the
Dirac phase factor is a well known feature of the electromagnetic Aharonov-Bohm
(AB) wave
function. In that case the solution for the Schr\"odinger equation for a
charged particle in the exterior of a solenoid containing a magnetic flux
$\phi$
can be constructed from the solution with no flux.
 One common exposition of  the AB effect is to consider two beams of
particles passing on opposite sides of the solenoid.
 The wave function for each
beam is constructed from the solution without magnetic flux using the
Dirac phase factor along each path.
The wave functions of the two beams will be related by  $\exp (i\oint A_i
dx^i)$,
which is precisely the holonomy of the electromagnetic
 $U(1)$ connection around the flux source.
One  then argues that the AB effect is due to the non triviality of the
holonomy even though the curvature of the connection $F_{\mu \nu}$ vanishes.
There is, however, a  problem with this reasoning:  the wave function
is not single valued.
This implies that, in spite of manifest axisymmetry, the angular momentum
of the particle is not conserved as one switches on the flux on the
solenoid adiabatically
 and is therefore not acceptable \cite{peshkin}.
Indeed, in the analysis of a scattering situation,
 studied by Aharonov and Bohm in detail, it is important that the wave
function is required to be single valued and the effect
is observed as a nontrivial scattering amplitude \cite{ab59}.
Thus, strictly speaking one cannot use the naive argument.
Nevertheless, one can still use the Dirac phase factor and
relate the AB effect to the holonomy of
a connection following Berry's
suggestion that the AB effect is a particular case of Berry's phase
 \cite{berry84}. However, the experimental setup has changed in this
setting: one considers
a particle contained in a box and the interference occurs between two
particles one of which was transported around the flux line.
Berry's phase is a particular case of the
Aharonov-Anandan geometric phase which we will now introduce \cite{aa87}.

Consider  a Hilbert space $\cal H$ and
the set ${\cal H}_o$ of normalized vectors, i.e.,

\be
{\cal H}_o = \{ | \Psi >\, \in {\cal H} \, / < \Psi | \Psi > = 1
\}\, .\label{6} \ee

Define an equivalence relation $\sim$ in ${\cal H}_o$ by:
 $|\Psi > \sim |\Phi >$
iff $|\Psi > = e^{i f} |\Phi >$ where $f$ is real.
The ray space $\cal P$ is defined as the quotient space,
\be
{\cal P} = {\cal H}_o / \sim \label{7}
\ee

\ni
and represents the space of all physically distinct states.
${\cal H}_o$ has the structure of a $U(1)$ principal bundle over $\cal P$.
Suppose that the system undergoes a cyclic evolution in $\cal P$
generated by a Hamiltonian $H$, $i\frac{d}{dt}|\Psi> = H(t)|\Psi>$.
 Denote by $C$ the corresponding curve,  $C:\ [0,\tau ]
\longrightarrow \cal P$.
The initial and final states of a curve $\hat{C}$ in ${\cal H}_o$
that is projected to $C$ in the ray space $\cal P$ will differ by a
phase,
\be
|\Psi (\tau )> = e^{i \beta} |\Psi (0) >.\label{8}
\ee
\ni
This phase $\beta$ can be decomposed into  two parts,
$\beta = \gamma + \delta$,
 where $\delta$ is called the {\it dynamical phase} which depends on the
Hamiltonian,
\be
\delta = - \int^\tau _0 <\Psi | H(t) \Psi >dt \, ,\label{9}
\ee
while the remainder, $\gamma$, is the {\it geometrical phase} which depends
only on
  the curve $C$ in $\cal P$.
There is a natural $U(1)$ connection defined in ${\cal H}_o$ such that the
holonomy around the path yields the geometric phase $\gamma$. Given $C$ in
$\cal P$,
let us define the horizontal lift $C'$ in ${\cal H}_o$,
by requiring that $|\Phi (t) >$ satisfies,
\be
<\Phi | \frac{d}{dt} \Phi >\, = 0\, .\label{10}
\ee
\ni
Then the state $|\Phi (t)>$ acquires a phase when parallel-transported:

\be
|\Phi (\tau) >\, = e^{ i \gamma} \ |\Phi (0) >\, .\label{11}
\ee

We can define a ``connection'' $A_s \equiv < \Psi (s) | \frac{d}{ds} \Psi
(s)>$.
Under ``gauge transformation'' $|\Psi > \longrightarrow e^{i \lambda (s)}
|\Psi >$, $A_s$ transforms as $A_s \longrightarrow A_s + \dot {\lambda}$.
Taking
$|\Psi '>$ such that $|\Psi '(\tau) > = |\Psi '(0) >$ the geometric phase is
\be
\gamma = i \oint <\Psi '| \dot {\Psi }' >\, .\label{12}
\ee

Berry's phase is recovered when the evolution in ${\cal H}_o$ is
 adiabatic and the time dependence of the Hamiltonian is completely contained
 on external  parameters $R_i(t)$,
 such that
 $R_i(\tau)= R_i (0)$.
 The system evolves
according to the Schr\"odinger equation
\be
i \, \frac{d}{dt} |\Psi (t)> = H(R_i(t))\,|\Psi (t)>\, .
\ee
The adiabaticity is imposed by assuming that an eigenstate $|E(R_i(0)) >$ will
 evolve with $H$ and will be in the state $|E(R_i(t)) >$ at time $t$.

The geometric phase is then
\be
\gamma (C) = i \, \oint_C\,<E(R_i)|\nabla _R E(R_i)> \,dR \, .\label{13}
\ee

In this framework, the Aharonov-Bohm effect can be described as follows
\cite{berry84}.
Consider a particle of charge $q$ inside a
``perfectly reflecting'' box such that the wave function $\Psi _n(r)$
 is non zero only
in the interior of the box. Call $R_i$ the vector from the origin of the
coordinate system (where the solenoid is located) to the center of the
box. When there is no magnetic potential
    the wave functions  have the form $\Psi _n(r-R)$ with energies $E_n$
 independent of $R_i$.  With non-zero flux the wave functions
$<\, r|\,n(R_i)>$ are
 obtained by the Dirac phase factor inside the box,
\be
<\,r|\,n(R_i)> = \exp \left[\, i q \int^r_R \,dr' \cdot A(r')
\right]\,\Psi _n (r-R) \, .
\ee
\ni
Let the box be transported around a circuit $C$ threaded by the flux line. The
geometric phase (\ref{13}) can be then computed using the fact that
\ba
<n(R_i)|\, \nabla _R n(R_i)> & = & \int \, d^3r\,\Psi ^*_n (r-R)\left[
-i q
 A(R) \Psi _n(r-R) + \nabla _R \Psi _n(r-R)\right] \nl
 & = &-i\,q\,A(R)\, ,
\ea
\ni
which implies
\be
\gamma(C) = q\,\oint\,A(R)\cdot dR = q\,\phi\, .
\ee
\ni
Then, the effect manifests itself as an  interference
between the particle in the
transported box and the one in a box which was not transported round
the circuit.

We conclude with  two remarks. First note that
the solution is single valued and therefore well defined
everywhere. Secondly, the factor $e^{i\gamma}$ coincides with the
holonomy of the electromagnetic
connection, but as we have seen the connection in question is the
 $U(1)$ connection on the unit sphere  ${\cal H}_o$.
 This fact   will allow us to relate the electromagnetic
and gravitational AB effects.

\section{Space of Solutions}

For a geometric phase as described in the last section to exist, the only
conditions
that must be satisfied are
the existence of a Hilbert Space with a Hermitian inner product and a unitary
 operator generating time evolution \footnote {More general conditions
 (i.e. non unitary evolution)
have been considered
in the literature \cite{sam}, but we will not consider them here.}.
The existence of such structures in the space-times considered in section 2 is
 the subject of this section.

In ordinary non-relativistic quantum mechanics, the time evolution of the
system is
governed by a differential equation which is first order in time, namely the
 Sch\"odinger equation.
 The
Hamiltonian operator is the infinitesimal generator of time translations and it
must be self-adjoint with respect to the Hermitian inner product in order to
generate unitary evolution. The Klein-Gordon equation is on the other hand a
{\it second order in time} equation and it is not clear a-priori that there
exist for general space-times a well defined ``square root'' that can be
identified
with the Hamiltonian.

The Hilbert space structure is however well defined in the case when the
underlying
 space-time is  stationary \cite{ash75b}
 and a Hamiltonian operator can be constructed as the generator
 of time translations. For completeness, let us recall this construction
where
the one-particle Hilbert space  $\cal H$ is obtained from the vector space
$V$ of real solutions to the Klein-Gordon equation.
 $\cal H$ must have the structure
of a complex Hilbert space in order to represent quantum states of a particle.
For that we need to introduce on $V$ a complex structure $J$
 and define, on the
complex vector space $(V,J)$, a Hermitian inner product $<\,|\,>$. Recall
that a complex structure $J$ is a linear map $j:V \rightarrow V$ such
that $J^2=-1$.

The one particle Hilbert space $\cal H$ would be the Cauchy
completion of the complex inner product space $(V,J,<|>)$.
 The complex structure must be compatible with the natural symplectic structure
$\Omega$ on $V$;
\be
\Omega (\Psi _1,\Psi _2) :=  \int _\Sigma dS^a
 (\Psi _2\nabla _a\Psi _1 -\Psi _1\nabla _a\Psi _2 )\, ,
\ee
where $\Sigma$ is any (3-dimensional) Cauchy surface on $(M,g^{ab})$.
\ni
$J$ is said to be compatible with $\Omega$ iff $\Omega (J\Phi _1,\Phi
_2)$ is a symmetric, positive definite metric $g(\cdot,\cdot)$ on $V$.

The Hermitian inner product is thus defined,
\be
<\Psi _1|\,\Psi _2> := \frac{1}{2} g(\Psi _1,\Psi _2) + \frac{1}{2} i
\Omega (\Psi_1,\Psi _2)\, .
\ee
In the case of stationary space-times the complex structure
 can be defined from the generator of {\it time evolution} ${\cal L}_t$ on
$V$ (${\cal L}_t$ is a well defined operator in
$V$ because it commutes with the Klein-Gordon operator). The complex
structure \be
J :=  - (- {\cal L}_t {\cal L}_t)^{-\frac{1}{2}}\cdot {\cal L}_t
\ee
\ni
satisfies all the required properties and therefore gives to $(V,J,<|>)$
the structure of a complex inner product space.
Given a complex structure $J$, one can recover the more familiar
 language of positive
and negative frequency decomposition. In this case the Hilbert Space $\cal H$
consists of positive frequency solutions $\Psi ^+$ of the Klein-Gordon
equation, \be
\Psi ^+ := \frac {1}{2} (\Psi - i\,J\,\Psi)\, ,
\ee
\ni
and the inner product takes the form,
 \be
<\Psi ^+_1|\Psi ^+_2> = -i\,\Omega (\overline{\Psi ^+_1},\Psi
^+_2)\, . \label{16} \ee

On this space the Schr\"odinger equation is given by
\be
H\cdot \Psi^+ = i \,\hbar \,{\cal L}_t \Psi^+
\ee

This covariant approach  (in which one
 takes real solutions of the KG equation in the full space-time to be
 the phase space)
 is completely equivalent to the ``canonical'' approach in which one
 considers the phase space to be pairs $(f,p)$ where $f$ is the initial wave
function on $\Sigma$: $f=\Psi |_\Sigma$, and $p$ is the initial
 normal derivative: $p=n^a \nabla_a \Psi |_\Sigma$.

\section{Geometric Phases in Gravity}

We can now investigate the gravitational Aharonov-Bohm
effect by combining the results that were established in the previous sections.
In
section 2, we found a solution to the Klein-Gordon equation outside a spinning
cosmic string. Let us now confine our quantum system to a box
situated at $R_i=(R_0,\theta_0,z_0)$ for $R_0> 4J/(1-4\mu)$. This procedure
allows us to get rid of two problems: the multivaluedness of the wave
function and having to deal with closed timelike curves
\footnote{ For $R_0 < \frac{4J}{(1-4 \mu )}$ the rotational Killing
vector field
$R^a = \left( \frac{\partial}{\partial \phi} \right)^a$ , which has closed
orbits, becomes timelike.}.

The boundary condition for the wave function inside the box is $\Psi
|_{\partial {\rm box}} = 0$. The Hilbert space structure for such solutions
can be  defined following the procedure outlined in section 4.

Following the electromagnetic case, we proceed by transporting the box
round a closed circuit $C$. Since the space-time is axisymetric we can
transport the box along orbits of the Killing vector field $R^a$. The
action of
 ${\cal L}_R$ commutes with  the Klein-Gordon operator $( \Box -
m^2)$. Consequently the action of transporting the box maps the
Hilbert space into itself.  Therefore we can restrict our analysis to the
Hilbert space  ${\cal H}_R$ for the box in the position $R$.
 Since we are considering the
covariant approach, the unitary motion generated by ${\cal L}_R$ in
 ${\cal H}_R$  takes the state $\Psi$ back to the same point in the
Hilbert space.

All the framework required to discuss Berry's phase is established.
We can now calculate the gravitational geometric phase using (\ref{12}).
We have
\be
\gamma = i \oint_C <\Psi (R_i)| {\cal L}_{R} \Psi(R_i)>\, ,\label{geop}
\ee
\ni
where $\Psi(R_i)$ is given by (\ref{5a}).

The integrand in (\ref{geop}) is evaluated using the Klein-Gordon inner
product (\ref{16}) as follows
 \ba
 <\Psi (R_i)| {\cal L}_{R} \Psi(R_i)> & = & i\int_\Sigma dS^a (\bar{\Psi}
\nabla_a({\cal L}_{R}\Psi) - ({\cal L}_{R}\Psi)\nabla_a
\bar{\Psi})\nonumber \\
& = & -i4 E J
\ea

\ni
The geometric phase (\ref{geop}) is then
\be
\gamma (C) =  8\, \pi\, J\, E \label{17}
\ee

The effect can be observed by interfering the wave-function in the
transported box and another that followed the orbits of the timelike
Killing vector field $t^a$.
We have avoided the region of closed timelike curves
by restricting the particle to the box and therefore we do not have
problems with the  non-Hermiticity of the Hamiltonian \cite{jack89}.

\section{Discussion}

Let's summarize the main results of the paper.

 We have found that for those stationary space-times whose Killing vector
field  has a  vanishing twist, solutions to the Klein-Gordon
equation can be found using the Dirac phase factor.
 A geometric phase can thus be found for those space-times given that
the existence of a Hermitian inner product in the Hilbert space of solutions
is assured.
 The construction of such phase and its comparison with the
electromagnetic AB effect
has been possible because we have interpreted both cases as the holonomy
of a $U(1)$ connection {\it on the space of states}. The phase depends on the
geometry of the closed path in the ray space, but will be nonzero
 when the topology of the underlying space-time is nontrivial. We can say,
therefore, that it  is of a topological origin.
For a spinning cosmic string, we have constructed the phase for a
particle going around the string.

We conclude with two remarks. In $2+1$ gravity, the form of the metric
 (\ref{4a}) for the
spinning particle is the general asymptotic form at infinity for
``asymptotically flat" space-times. Therefore, the effect is
always present near infinity for nontrivial topologies ($\Pi _1(\Sigma)\neq
0$), when the total angular momentum is non vanishing ($J\neq 0$).
Secondly,
our result is a generalization of the geometric phase for a Klein Gordon
particle in Minkowski space-time studied in \cite{amaz93}.

\vskip.5cm
\ni
{\Large\bf  Acknowledgments}
\vskip.3cm

We would like to thank A. Ashtekar for suggesting this problem. The
completion of
this work would not have been possible without his advise and extensive
discussions.
We also thank D. Marolf,  M. Varadarajan and J.A. Zapata
for discussions. AC was supported by the National University
of M\'exico (DGAPA-UNAM). MP was supported by CAPES.
This work was supported in part by the NSF grant PHY93-96246 and by the Eberly
research fund of Penn State University.

\end{document}